\documentclass[twocolumn,floatfix,superscriptaddress,a4paper,showkeys,nofootinbib,reprint]{revtex4-1}
\usepackage{epsfig}
\usepackage{latexsym}
\usepackage{xspace}
\usepackage[colorlinks=true,linktocpage=true,linkcolor=blue,citecolor=blue,allcolors=blue]{hyperref}
\usepackage[utf8]{inputenc}
\usepackage{indentfirst}
\usepackage{enumerate}
\usepackage{color}

\usepackage[caption=false,position=top]{subfig}

\usepackage{amsmath}
\usepackage{amssymb}
\usepackage[english]{babel}
\usepackage{url}

\newcommand{\eq}[1]{\begin{align} #1 \end{align}}

\begin{document}

\title{
Quantum van der Waals theory meets quarkyonic matter
}

\author{Roman~V.~Poberezhnyuk}
\affiliation{Bogolyubov Institute for Theoretical Physics, 03680 Kyiv, Ukraine}
\affiliation{Frankfurt Institute for Advanced Studies, Giersch Science Center, Ruth-Moufang-Str. 1, D-60438 Frankfurt am Main, Germany}

\author{Horst~Stoecker}
   \affiliation{Frankfurt Institute for Advanced Studies, Giersch Science Center, Ruth-Moufang-Str. 1, D-60438 Frankfurt am Main, Germany}
    \affiliation{
GSI Helmholtzzentrum f\"ur Schwerionenforschung GmbH, Planckstr. 1, D-64291 Darmstadt, Germany}
    \affiliation{Institute of Theoretical Physics, Goethe Universit\"at, Frankfurt, Germany}  

\author{Volodymyr~Vovchenko}
\affiliation{Physics Department, University of Houston, Box 351550, Houston, TX 77204, USA}
\affiliation{Frankfurt Institute for Advanced Studies, Giersch Science Center, Ruth-Moufang-Str. 1, D-60438 Frankfurt am Main, Germany}

\begin{abstract}
We incorporate the empirical low-density properties of isospin symmetric nuclear matter into the excluded-volume model for quarkyonic matter by 
including attractive mean field in the nucleonic sector and considering variations on the nucleon excluded volume mechanism. 
This corresponds to the quantum van der Waals equation for nucleons, with the interaction parameters fixed to empirical ground state properties of nuclear matter. 
The resulting equation of state exhibits the nuclear liquid-gas transition at $n_B \leq \rho_0$ and undergoes a transition to quarkyonic matter at densities $n_B \sim 1.5-2 \rho_0$ that are reachable in intermediate energy heavy-ion collisions. The transition is accompanied by a peak in the sound velocity. The results depend only mildly on the chosen excluded volume mechanism but do require the introduction of an infrared regulator $\Lambda$ to avoid the acausal sound velocity.
We also consider the recently proposed baryquark matter scenario for the realization of the Pauli exclusion principle, which yields a similar equation of state and turns out to be energetically favored in all the considered setups.
\end{abstract}


\keywords{quarkyonic matter, baryquark matter, nuclear liquid-gas transition, quark-hadron duality}

\maketitle

\section{Introduction}

Cold nuclear matter is expected to undergo a transition from hadrons to quarks as the baryon density is increased.
Neutron star observations indicate a soft-hard evolution of the neutron-rich nuclear equation of state at zero temperature, with a peak in the sound velocity exceeding the conformal limit~\cite{Tews:2018kmu,Fujimoto:2019hxv,Tan:2020ics,Altiparmak:2022bke}.
The concept of quarkyonic matter~\cite{McLerran:2007qj,McLerran:2018hbz} 
provides a promising mechanism describing this phenomenon.
Motivated by the anticipated QCD features 
in the large $N_c$ limit~\cite{McLerran:2007qj,Philipsen:2019qqm,Kovensky:2020xif},
it envisions the existence of a quark-baryon mixture with a mixed phase in momentum space.
Here deconfined quarks fill the Fermi sea while baryonic excitations exist around the Fermi surface only.
The peak in the sound velocity is associated with the transition between nucleon dominated and quarkyonic regimes.

In quarkyonic matter, the dynamical generation of the momentum shell structure proceeds through energy minimization at fixed baryon density.
This requires either the presence of strong nuclear repulsion~\cite{Jeong:2019lhv} or quark interactions~\cite{Duarte:2023cki}, or both.
Excluded volume repulsion, originally developed for nuclear matter~\cite{Rischke:1991ke}, is one such mechanism and will be used throughout this work.
Various generalizations of the theory~\cite{Sen:2020peq,Duarte:2020xsp,Zhao:2020dvu,Cao:2020byn,Duarte:2020kvi,Duarte:2021tsx} have been developed.
Quarkyonic matter implementations have also been considered for the description of neutron stars, see e.g. Refs.~\cite{McLerran:2018hbz,Han:2019bub,Sen:2020qcd}.

Recently, baryquark matter was proposed~\cite{Koch:2022act}, which is similar to quarkyonic matter; however, it assumes the opposite momentum shell structure with nucleons occupying low momentum states 
and surrounded by a shell of quarks.
It was shown that baryquark matter is energetically favored relative to quarkyonic matter in existing implementations and, moreover, it yields a physically acceptable behavior of the speed of sound without the need to introduce an infrared regulator.
We, therefore, consider the baryquark matter configuration in addition to quarkyonic matter in all our calculations.

All excluded-volume based 
frameworks for quarkyonic (and baryquark) matter studied so far neglect a realistic description of the low-density nuclear equation of state and the associated nuclear liquid-gas transition. As such, 
the quark onset density and the associated peak in the speed of sound 
were determined by a free parameter -- the nucleon excluded volume.
In the present work, we address this shortcoming by introducing an attractive mean-field corresponding to the quantum van der Waals equation of Ref.~\cite{Vovchenko:2015vxa} and fixing the interaction parameters to reproduce the properties of the nuclear ground state.
We also consider variations on the 
excluded volume mechanism~\cite{Vovchenko:2017cbu}.
Implementing these normal nuclear matter constraints, we find that the peak in the speed of sound occurs at $1.5-2$ times the normal nuclear density, in both quarkyonic and baryquark matter descriptions.
Our work illustrates the influence of normal nuclear matter properties on the realization of the quark-hadron duality at zero temperature.
Previously, it has been shown that the same nuclear matter constraints affect the quark-hadron crossover region at finite temperature and moderate densities, including the susceptibilities of conserved charges~\cite{Vovchenko:2016rkn,Mukherjee:2016nhb,Motornenko:2019arp}, chemical freeze-out in heavy-ion collisions~\cite{Poberezhnyuk:2019pxs}, and the convergence radius of Taylor expansion around $\mu_B = 0$~\cite{Savchuk:2019yxl}.

The paper is organized as follows.
Section \ref{sec:Pauli} briefly describes the quasiparticle mixture of nucleons and quarks, as well as the different considered realizations of Pauli exclusion principle acting between them.
Section~\ref{sec:interaction} implements different scenarios for interactions between nucleons.
Section~\ref{sec:results} presents the resulting equations of state of quarkyonic and baryquark matter.
Summary and discussion in Sec.~\ref{sec:summary} close the article.

\begin{figure*}
    \includegraphics[width=0.99\textwidth]{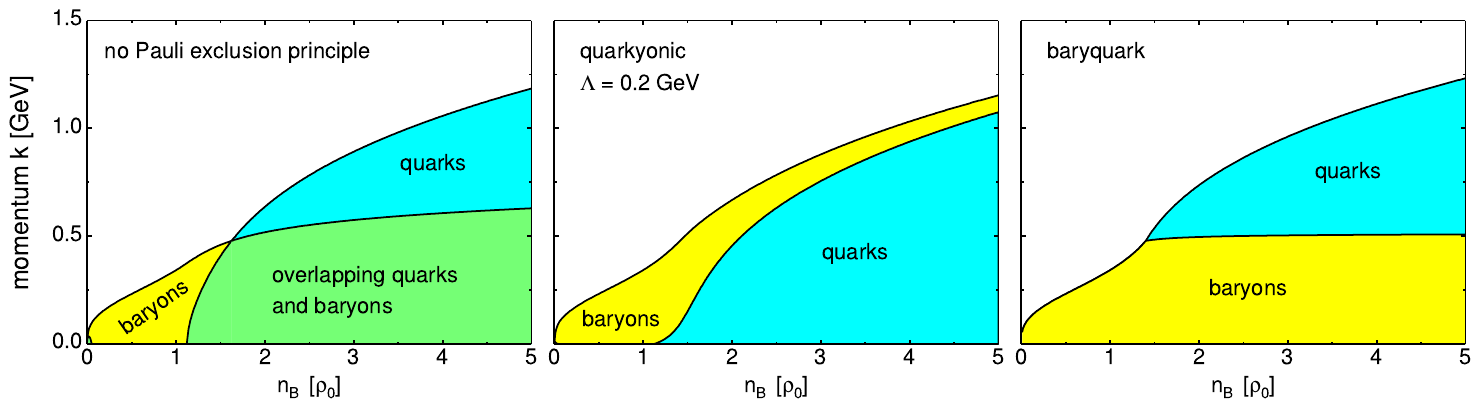}
    \caption{Density dependence of baryon and quark composition in momentum space for different realizations of the Pauli exclusion principle acting between baryons and quarks.
    Nuclear interactions are modeled by the quantum van der Waals equation.
    The momenta of quarks are multiplied by $N_c$. The baryon density scale is in units of normal nuclear density $\rho_0$.
    }
    \label{fig:shell-structures}
\end{figure*}

\section{Nucleons and quarks with Pauli exclusion principle}
\label{sec:Pauli}

Let us first consider a mixture of non-interacting baryons and quarks.
In the following, we will denote the momentum of baryons by $k$ and the momentum of quarks by $q$.
At zero temperature, $T=0$, all baryons and quarks occupy all the momentum states up to the Fermi levels $k_F$ and $q_F$, respectively. 
The densities of baryon number and energy contain contributions from quarks and baryons are
$n_B=n_{Q}^{\rm id}(q_F)+n^{\rm id}_N(k_F)$ and $\varepsilon=\varepsilon_Q^{\rm id}(q_F)+\varepsilon^{\rm id}_N(k_F)$. 
The ideal gas functions read
\eq{\label{eq:nN}
n_N^{\rm id}(k_F)&=\frac{g}{2\pi^2}\int_{0}^{k_F}k^2 \, \rho_N(k) dk, \\
\label{eq:eN}
\varepsilon_N^{\rm id}(k_F)&=\frac{g}{2\pi^2}\int_{0}^{k_F}k^2 \, \rho_N (k) \, \epsilon_N(k) dk, \\\label{eq:nq}
n_Q^{\rm id}(q_F)&=\frac{g}{2\pi^2}\int_{0}^{q_F}q^2 \, \rho_Q(q) dq, \\\label{eq:eq}
\varepsilon_Q^{\rm id}(q_F)&=N_c\frac{g}{2\pi^2}\int_{0}^{q_F} q^2 \, \rho_Q(q) \, \epsilon_Q(q) dq.
}
Here $\rho_{Q}(k),\rho_{N}(q)$ are the modification factors for the phase space densities of quark and baryon states due to the Pauli exclusion principle (explained below),
$\epsilon_{Q}(q)= \sqrt{m_{Q}^2+q^2}$ and
$\epsilon_{N}(k)= \sqrt{m_{N}^2+k^2}$ (with $m_N = N_c m_Q$) 
are relativistic single-particle energies,
$g = g_Q = g_N = 4$ is the spin-isospin degeneracy factor for quarks and nucleons.
Note that the quarks have an additional degeneracy factor of $N_c$ due to color.
This factor is canceled out by the $1/N_c$ fractional baryon charge in the expression~\eqref{eq:nq} for quark contribution to the baryon density but is present in Eq.~\eqref{eq:eq} describing quark energy density.

If one regards the confined and deconfined quarks 
as independent fermions, then these two different particles can occupy the same quantum state, i.e. the Pauli exclusion principle does not apply. 
The coexistence phase is the mixture of the quarks and baryons, i.e. $\rho_{N,Q}=1$.
The values of $q_F$ and $k_F$ would correspond to the configuration of the minimized energy density at a fixed value of the baryon number density $n_B$.
The possible baryon ($k_F$) and quark ($N_c q_F$) Fermi momentum dependencies on the baryon density $n_B$ are shown in the left panel of Fig.~\ref{fig:shell-structures}.
One can see a significant overlap between the quark and baryon momentum states.

In a more realistic setup, one must account for the Pauli exclusion principle between the confined and deconfined quarks, namely that they cannot both occupy the same state~\cite{McLerran:2018hbz,Jeong:2019lhv}.\footnote{Note, that in quarkyonic matter, the color confinement persist till large densities which substantiates the realization of the Pauli exclusion principle. Thus, the terms "confined" and "deconfined quarks" are used conditionally to distinguish between the two types of quasiparticles.}
We shall consider here a simplified picture, where we neglect the internal motion of quarks inside a baryon, thus, each confined quark carries exactly $1/N_c$ fraction of nucleon momentum and the relation between the Fermi momenta $q_F$ of quarks and $k_F$ of nucleons reads $q_F = k_F / N_c$.
Then, a deconfined quark with a given momentum $q$ and spin-isospin projection Pauli-blocks a possible nucleon state with momentum $k = N_c q$ that would otherwise contribute a confined quark with momentum $q$ and the same spin-isospin quantum numbers.

In symmetric nuclear matter that we consider here, the Pauli principle implies the following condition which relates the quark and nucleon densities of states entering Eqs.~\eqref{eq:nN}-\eqref{eq:eN} for all $k < k_F$:
$\rho_N(k) = 1 - \rho_Q(q)$ with $q = k/N_c$.
It is seen that the presence of deconfined quarks leads to a reduction of nucleon density of states at an appropriately scaled momentum.
The additional suppression of nucleon density of states could also emerge by incorporating the quark substructure effects, for instance through explicit duality between quark and baryon distributions as was recently done in Ref.~\cite{Fujimoto:2023ioi}.
Although we do not explicitly incorporate the quark substructure effects in the present work, this additional suppression is effectively present in our model through baryon excluded volume effect described in the next section.

To complete the description it remains to define the momentum dependence of the quark density of states $\rho_Q(q)$.
We consider two simple forms for $\rho(k)$ which correspond to two opposite scenarios for the realization of the Pauli exclusion principle.

\subsection{Quarkyonic matter}
\label{sec:Pauli:qy}

Baryonic excitations around the Fermi surface are the hallmark feature of quarkyonic matter. 
It is motivated by the notion that small relative momentum interactions associated with confinement are blocked in the "bulk" Fermi sea of quarks, up to the momentum $q_{\rm bu} = k_{\rm bu} / N_c$. Thus, as density increases, confinement can occur only at the Fermi surface, leading to the dynamical generation of the shell of baryons with a thickness $\Delta = k_F - k_{\rm bu}$.
A practical realization of quarkyonic matter corresponds to a strict separation in momentum space where deconfined quarks occupy the low momentum states, $q < k_{\rm bu} / N_c$, whereas baryons occupy a shell with momenta lying between $k_{\rm bu} < k < k_F$~\cite{McLerran:2018hbz,Jeong:2019lhv}.
Such a realization of the Pauli exclusion principle corresponds to the following quark density of states 
\eq{
\rho_Q(q) = \Theta(q_{\rm bu}-q),
}
where $q_{\rm bu}= k_{\rm bu}/N_c$ and, consequently,
\eq{
\rho_N(k) = \Theta(k - k_{\rm bu}) \, \Theta(k_F - k).
}
This leads to the following explicit expressions for nucleon and quark contributions to the baryon and energy densities:

\eq{\label{eq:nN:qm}
n_N^{\rm id}(k_F)&=\frac{g}{2\pi^2}\int_{k_{\rm bu}}^{k_F}k^2 \,  dk, \\
\label{eq:eN:qm}
\varepsilon_N^{\rm id}(k_F)&=\frac{g}{2\pi^2}\int_{k_{\rm bu}}^{k_F}k^2 \, \epsilon_N(k) dk, \\\label{eq:nq:qm}
n_Q^{\rm id}(q_F)&=\frac{g}{2\pi^2}\int_{0}^{k_{\rm bu}/N_c}q^2 \,  dq, \\\label{eq:eq:qm}
\varepsilon_Q^{\rm id}(q_F)&=N_c\frac{g}{2\pi^2}\int_{0}^{k_{\rm bu}/N_c} q^2 \,  \epsilon_Q(q) dq.
}

The values of $k_F$ and $k_{\rm bu}$ at given baryon density are determined from minimizing energy density at fixed baryon density $n_B$.

Note that in this model the speed of sound exhibits singular behavior at the onset of quark appearance, as first investigated in Ref.~\cite{Jeong:2019lhv}.
As a result, adjustments to the quark sector of the model are needed to achieve a physically plausible behavior.
For example, introducing an infrared regulator $\Lambda$~\cite{Jeong:2019lhv} by multiplying $\rho_Q(q)\rightarrow \rho_Q(q)\frac{\sqrt{q^2+\Lambda^2}}{q}$ in the quark density of states in Eqs.~(\ref{eq:nq})-(\ref{eq:eq}). 
It is known, however, that such a modification of the density of states leads to an unrealistic feature of quarks reappearing at the lowest baryon densities.\footnote{This problem was addressed in Ref.~\cite{Sen:2020qcd} where a density-dependent infrared regulator $\Lambda(n_N)$ was introduced.}

The evolution of the resulting quarkyonic matter shell structure with baryon density is shown in the middle panel of Fig.~\ref{fig:shell-structures}.

\subsection{Baryquark matter}
\label{sec:Pauli:bq}

An opposite scenario to quarkyonic matter -- the baryquark matter -- was introduced in Ref.~\cite{Koch:2022act}.
Here the low momentum states are saturated by baryons and  they are surrounded by a Fermi shell of quarks. 
This structure is motivated by being more energetically favorable in the quasi-particle approach and, presumably, in transport simulations.
The quark density of states reads
\eq{
\rho_Q(q) = \Theta(q - q_{\rm bu}) \, \Theta(q_F - q)
}
and, consequently,
\eq{
\rho_N(k) = \Theta(k_{\rm bu}-k),
}
leading to
\eq{\label{eq:nN:bq}
n_N^{\rm id}(k_F)&=\frac{g}{2\pi^2}\int_{0}^{k_{\rm bu}}k^2 \,  dk, \\
\label{eq:eN:bq}
\varepsilon_N^{\rm id}(k_F)&=\frac{g}{2\pi^2}\int_{0}^{k_{\rm bu}}k^2 \, \epsilon_N(k) dk, \\\label{eq:nq:bq}
n_Q^{\rm id}(q_F)&=\frac{g}{2\pi^2}\int_{k_{\rm bu}/N_c}^{k_{\rm F}/N_c}q^2 \,  dq, \\\label{eq:eq:bq}
\varepsilon_Q^{\rm id}(q_F)&=N_c\frac{g}{2\pi^2}\int_{k_{\rm bu}/N_c}^{k_{\rm F}/N_c} q^2 \,  \epsilon_Q(q) dq.
}

The baryquark matter shell structure is shown in the right panel of Fig.~\ref{fig:shell-structures}.

\subsection{Other possibilities}

The two scenarios above can be considered extreme opposites among all possible realizations of the Pauli exclusion principle.
In Appendix~\ref{sec-uniform} 
we illustrate this by studying the 
uniform mixture of quarks and baryons which is homogeneous in both coordinate and momentum space and assumes a common Fermi surface for confined and deconfined quarks, i.e. $\rho_Q(q) = {\rm const}$.
A similar framework was used in Ref.~\cite{Stoecker:1980uk} where a quark-hadron mixture without separation in the momentum space was studied in the context of heavy-ion collisions.
More elaborate forms of $\rho_Q(q)$ may be considered as well.
For example, interactions between quarks lead to diffusion of the Fermi
surface between baryons and quarks~\cite{Glozman:2011eu}.
Another possibility is the existence of a more complex shell structure with multiple boundaries between quarks and baryons in the momentum space.

\section{Nuclear interactions}
\label{sec:interaction}

The introduction of interactions is necessary to obtain a dynamically generated appearance of quarks at sufficiently high baryon densities.
This is so because the energy density of a free nucleon gas is always smaller than that of a free quark gas at the same value of baryon number density, as originally pointed out in~\cite{Jeong:2019lhv}.
Following earlier works on the subject~\cite{Jeong:2019lhv,Sen:2020peq,Duarte:2021tsx,Koch:2022act}, we treat quark quasiparticles in the framework of an ideal gas,  $n_Q(q_F)=n_Q^{\rm id}(q_F)$. We also neglect quark-nucleon interactions apart from the Pauli exclusion principle described in the previous section.

Regarding the nuclear sector, it has been shown~\cite{Jeong:2019lhv} that the presence of strong nuclear repulsion, such as given by the excluded volume model~\cite{Rischke:1991ke}, makes pure nucleon system energetically unfavorable at large baryon number densities and thus gives rise to the emergence of quarkyonic matter above some threshold baryon density. 
In the excluded volume model the total available volume for nucleons to move in is reduced by subtracting their eigenvolumes, effectively giving rise to repulsive interactions.
It should be noted that the excluded volume repulsion is different from mean-field approaches to nuclear interactions in that the interactions is not defined locally from a Lagrangian and it (i) imposes a limiting maximum nucleon number density and (ii) modifies the relation between the nucleon's Fermi momentum and number density.
The resulting excluded volume model for quarkyonic matter has since been studied in a number of works~\cite{Duarte:2020xsp,Duarte:2020kvi,Sen:2020qcd,Duarte:2021tsx,Koch:2022act}.

One drawback of the excluded volume model for quarkyonic matter is that it does not reproduce the empirical properties of nuclear matter at normal nuclear densities. 
As the excluded volume interactions are purely repulsive, the model does not predict the existence of the nuclear ground state at any baryon density. 
Furthermore, the excluded volume parameter $b$, which largely determines at which baryon density the onset of quarks appears~\cite{Jeong:2019lhv}, has essentially been treated as a free parameter.\footnote{The excluded volume $b$ is related to the limiting nucleon density $n_0$ used in the notation throughout Refs.~\cite{Duarte:2020xsp,Duarte:2020kvi,Sen:2020qcd,Duarte:2021tsx,Koch:2022act} as $b \equiv 1/n_0$.}

These issues can be mitigated by generalizing the model to incorporate the basic properties of nuclear matter at normal nuclear densities.
First, one must incorporate attractive nuclear interactions to reproduce the nuclear ground state. The simplest way to do that is by
an attractive mean-field, resulting in the quantum van der Waals (QvdW) model of nuclear matter introduced in Ref.~\cite{Vovchenko:2015vxa}.
The QvdW interaction parameters $a$ and $b$ are then fixed to reproduce the nuclear ground state at $n_B = \rho_0 = 0.16$~fm$^{-3}$.
Further improvement is achieved by generalizations 
of the excluded volume mechanism~(beyond van der Waals) and incorporating non-linear density dependence of the mean-field.
This gives rise to the quantum real gas model~\cite{Vovchenko:2017cbu}~(or density-dependent van der Waals~\cite{Dutra:2020qsn}) of nuclear matter. 

Here we use the quantum real gas formalism of Ref.~\cite{Vovchenko:2017cbu} for the nucleonic sector in quarkyonic and baryquark matter.
The nucleonic number and energy densities are given by the following relations:
\eq{
\label{eq:nNvdw}
n_N & =f(n_N)~n_N^{\rm id}(k_F),\\
\label{eq:eNvdw}
\varepsilon_N &= f(n_N)~\varepsilon_N^{\rm id}(k_F)+ n_N \, u(n_N),
}
where $n_N^{\rm id}(k_F)$ and $\varepsilon_N^{\rm id}(k_F)$ are evaluated through Eqs.~\eqref{eq:nN} and~\eqref{eq:eN}, respectively, taking into account the Pauli exclusion principle.
Here $u(n_N)$ is the nucleon density dependent mean-field.
In this work, we limit our consideration to the attractive mean-field of the van der Waals model, which corresponds to
\eq{\label{eq:uN}
u(n_N) = -a n_N.
}

\begin{center}
\begin{table*}
\begin{tabular}
{ l | c | c | c | c  }
 & ~vdW~  & ~CS~  & ~TVM~  &~Experiment~ \\
\hline
\hline
~$a~[\rm{MeV} ~ fm^3]$               & 329 & 347 & 349 & --\\
~$b~[\rm fm^3]$               & 3.42   & 4.43 & 4.28 & --\\
~$\rho_0~[\rm fm^{-3}]$            & 0.16 & 0.16 & 0.16 & $0.15 \pm 0.01$~\cite{Horowitz:2020evx} \\
~$K_0$   [\rm{MeV}]                 & 763   & 601 & 564 & 250~-~315~\cite{Stone:2014wza} \\
 ~$T_c~[\rm{MeV}]$               & 19.7   & 18.6 & 18.3 & $17.9~\pm~0.4$~\cite{Elliott:2013pna}\\
 ~$n_c~[\rm{fm}^{-3}]$           & 0.072  & 0.070 & 0.069 & $0.06~\pm~0.01$~\cite{Elliott:2013pna}\\
\end{tabular}
\caption{
\label{tab:NM}
Values of the vdW interactions parameters, the resulting ground-state nuclear density, incompressibility, and the nuclear liquid-gas critical point location within the QvdW model with van der Waals~(vdW), Carnahan-Starling~(CS), and trivirial model~(TVM) excluded volume prescriptions.
}
\end{table*}
\end{center}

The function $f(n_N)$ in Eqs.~\eqref{eq:nN} and~\eqref{eq:eN} quantifies the fraction of available volume for nucleons to move in due to the excluded volume effect, thus $0 \leq f(n_N) \leq 1$.
We consider three different prescriptions for the excluded volume mechanism.

\begin{enumerate}
    \item {\it van der Waals (vdW) excluded volume}\\

    In the commonly used excluded volume model a~l\'a van der Waals, one simply subtracts the eigenvolumes of all nucleons which are assumed to be constant for all nucleons and independent of the density, corresponding to
    \eq{\label{eq:fvdw}
    f_{\rm vdW}(n_N)=1~-~bn_N.
    }
    This expression implies the existence of a limiting nucleon density $n_{\rm vdW}^{\rm lim} = 1/b$ at which the available volume vanishes, $f_{\rm vdW}(n_{\rm vdW}^{\rm lim}) = 0$.

    \item {\it Carnahan-Starling~(CS) excluded volume}\\

    In a generalized excluded volume mechanism, the relation between $f(n_N)$ and $n_N$ need not be linear.
    One well-known generalization is the Carnahan-Starling equation~\cite{carnahan1969equation}, which gives a remarkably accurate description of the equation of state for classical hard spheres over broad range of densities. The Carnahan-Starling model corresponds to the following choice of $f(n_N)$~\cite{Vovchenko:2017cbu}:
    \eq{\label{eq:fCS}
    f_{\rm CS}(n_N)=\exp\left[-\frac{3 b n_N}{4 - b n_N}-\frac{4 b n_N}{(4 - b n_N)^2}\right]~.
    }

    At small densities, $bn \ll 1$, the CS model reduces to the vdW model, i.e. $f_{\rm CS} = f_{\rm vdW} + O[(b n_N)^2]$.
    However, the behavior is different at large densities. 
    In particular, the limiting nucleon density, which corresponds to $f(n^{\rm lim}) = 0$, is factor four larger in the CS model than in the vdW model: $n_{\rm CS}^{\rm lim} = 4 n_{\rm vdW}^{\rm lim} = 4/b$.

    The CS model can be regarded as the vdW model with density-dependent effective excluded volume parameter
    \eq{\label{eq:beff}
    b^{\rm eff}(n_N) = \frac{1-f(n_N)}{n_N}.
    }
    In the vdW model, the effective excluded volume per Eq.~\eqref{eq:beff} is a constant $b_{\rm vdW}^{\rm eff} = b$ at all densities. In the CS model, on the other hand, the effective excluded volume is a decreasing function of the nucleon density, going from $b_{\rm CS}^{\rm eff} \simeq b$ at small densities $n_N \ll 1/b$ down to $b_{\rm CS}^{\rm eff} \simeq b/4$ near the limiting density $n_{\rm CS}^{\rm lim} = 4/b$.

    The CS model yields a softer equation of state compared to the standard vdW excluded volume, and can extend the causality range of the dense hadronic matter EoS. Recently, it has been successfully applied to study neutron star matter within hadron resonance gas framework~\cite{Fujimoto:2021dvn}.

    \item {\it trivirial model~(TVM) excluded volume}\\

    Our final consideration is the trivial model~(TVM), originally introduced in Ref.~\cite{Vovchenko:2019hbc} to study the analytic properties of the partition function associated with the liquid-gas transition.
    The available volume fraction reads
    \eq{\label{eq:fTVM}
    f_{\rm TVM}(n_N)= \exp\left[-bn_N-\frac{b^2 n_N^2}{2}\right].
    }

    The TVM model, in contrast to vdW and CS models, has no limiting density, that is $f(n_N) > 0$ for any finite $n_N$.
    Like the CS model, it reduces to the vdW model in the low-density limit, $b n_N \ll 1$. The effective excluded volume parameter $b^{\rm eff}_{\rm TVM}(n_N)$ monotonically decreases with nucleon density $n_N$ from $b_{\rm TVM}^{\rm eff} \simeq b$ at small densities, $n_N \ll 1/b$, toward zero, $b_{\rm TVM}^{\rm eff} \to 0$, for $n_N \to \infty$.
    
\end{enumerate}

The parameters $a$ and $b$ are fixed in each of the three models to reproduce the ground state of symmetric nuclear matter at $n_N = \rho_0 = 0.16$~fm$^{-3}$~(see Refs.~\cite{Vovchenko:2015vxa,Vovchenko:2017cbu,Vovchenko:2019hbc} for the details of the procedure).
Table~\ref{tab:NM} lists the resulting values of the interaction parameters $a$ and $b$, as well as the predicted ground state incompressibility $K_0$, and the location of the nuclear liquid-gas critical point, along with the available experimental estimates.
It is seen that the models do a decent job in describing the critical point location but overestimate considerably the empirical estimates for nuclear incompressibility factor $K_0$.
The density-dependent excluded volume prescriptions studied here improve the description of $K_0$ somewhat, but not sufficiently enough to reproduce the empirical constraints. 
This issue can be addressed by considering density-dependent mean fields~\cite{Vovchenko:2017cbu,Dutra:2020qsn}.
We leave these improvements of nuclear matter description for future work and instead focus on the transition to quarkyonic/baryquark matter.

\section{Quantum van der Waals quarkyonic and baryquark matter}
\label{sec:results}

Having defined the interactions between nucleons we can now return to the topic of the equation of state of quark-nucleon mixture.
The baryon number and energy densities contain contributions from quarks and nucleons,
\eq{
n_B(k_F,k_{\rm bu}) & = n_Q(q_F;q_{\rm bu}) + n_N(k_F;k_{\rm bu}), \\
\varepsilon(k_F,k_{\rm bu}) & = \varepsilon_Q(q_F;q_{\rm bu}) + \varepsilon_N(k_F;k_{\rm bu}).
}
Here $q_F = k_F/N_c$, $q_{\rm bu} = k_{\rm bu}/N_c$, and $n_{B,Q}$ and $\varepsilon_{B,Q}$ are defined by Eqs.~(\ref{eq:nN})-(\ref{eq:eq}).

In both quarkyonic and baryquark matter setups, the thermodynamic functions depend on two parameters, the Fermi momentum $k_F$, and the ``bulk'' Fermi sea momentum $k_{\rm bu}$ that delineates the boundary between the confined and deconfined quarks in the momentum space.
These two parameters define the value of baryon number density $n_B$, and the quark fraction $n_Q/n_B$.
The same is true in the opposite direction: the values of $n_B$ and $n_Q/n_B$ unambiguously define the value of $k_F$ and $k_{\rm bu}$.
We use this fact to calculate the equation of state by minimizing the energy density $\varepsilon$ at fixed baryon number density $n_B$ with respect to the quark fraction $n_Q/n_B$.

\begin{figure*}
\includegraphics[width=.32\textwidth]{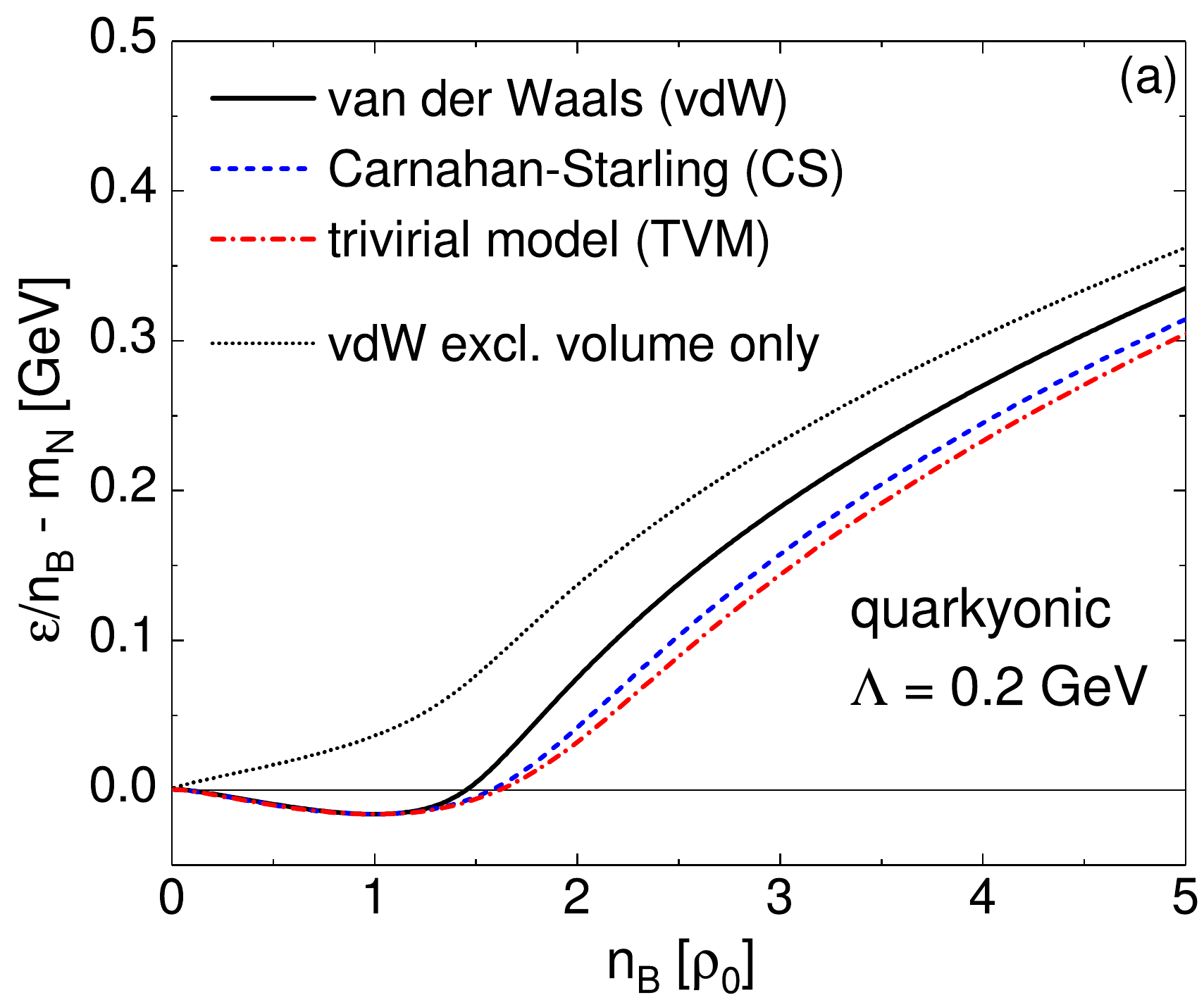}
    \includegraphics[width=.32\textwidth]{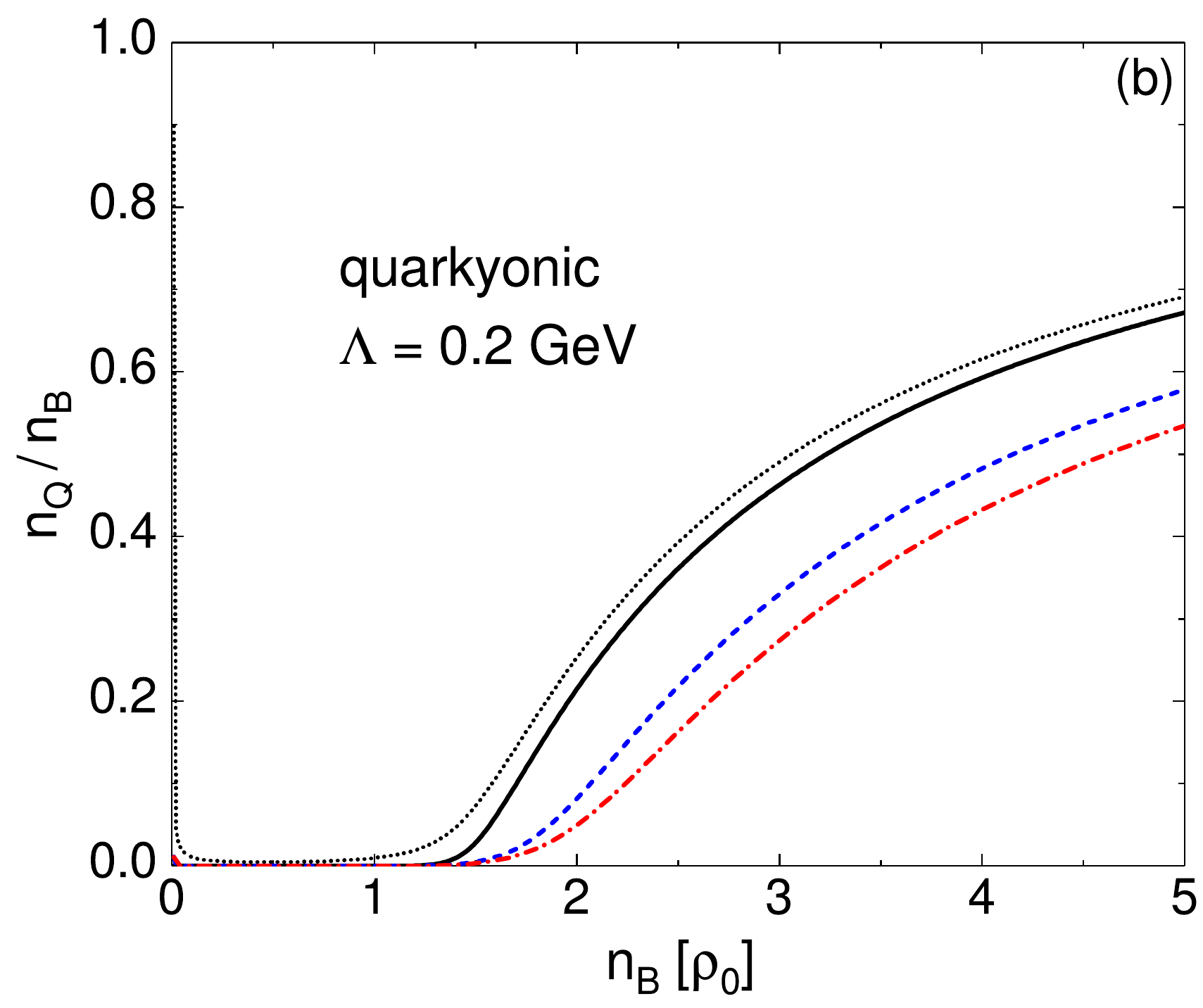}
     \includegraphics[width=.32\textwidth]{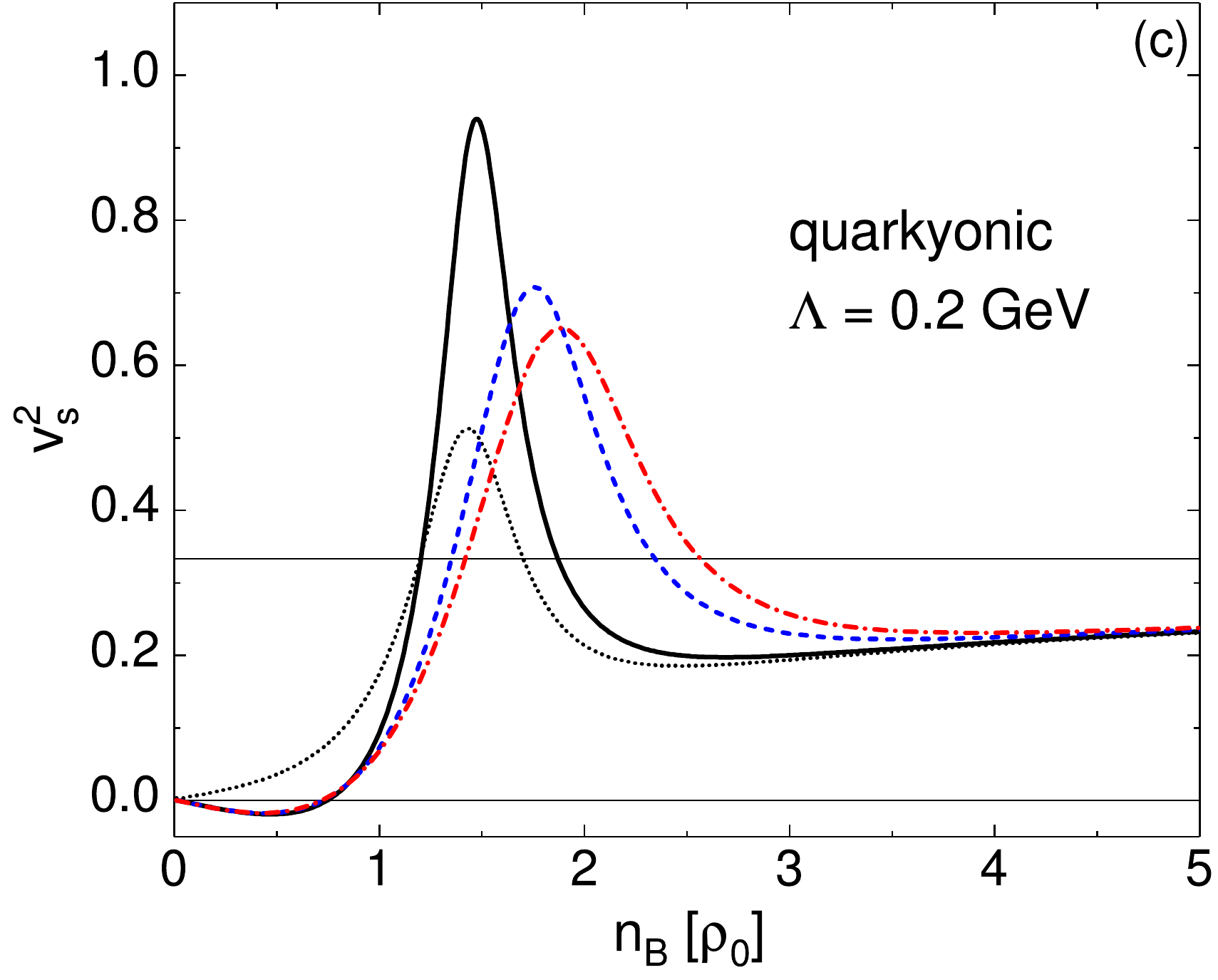}
     \includegraphics[width=.32\textwidth]{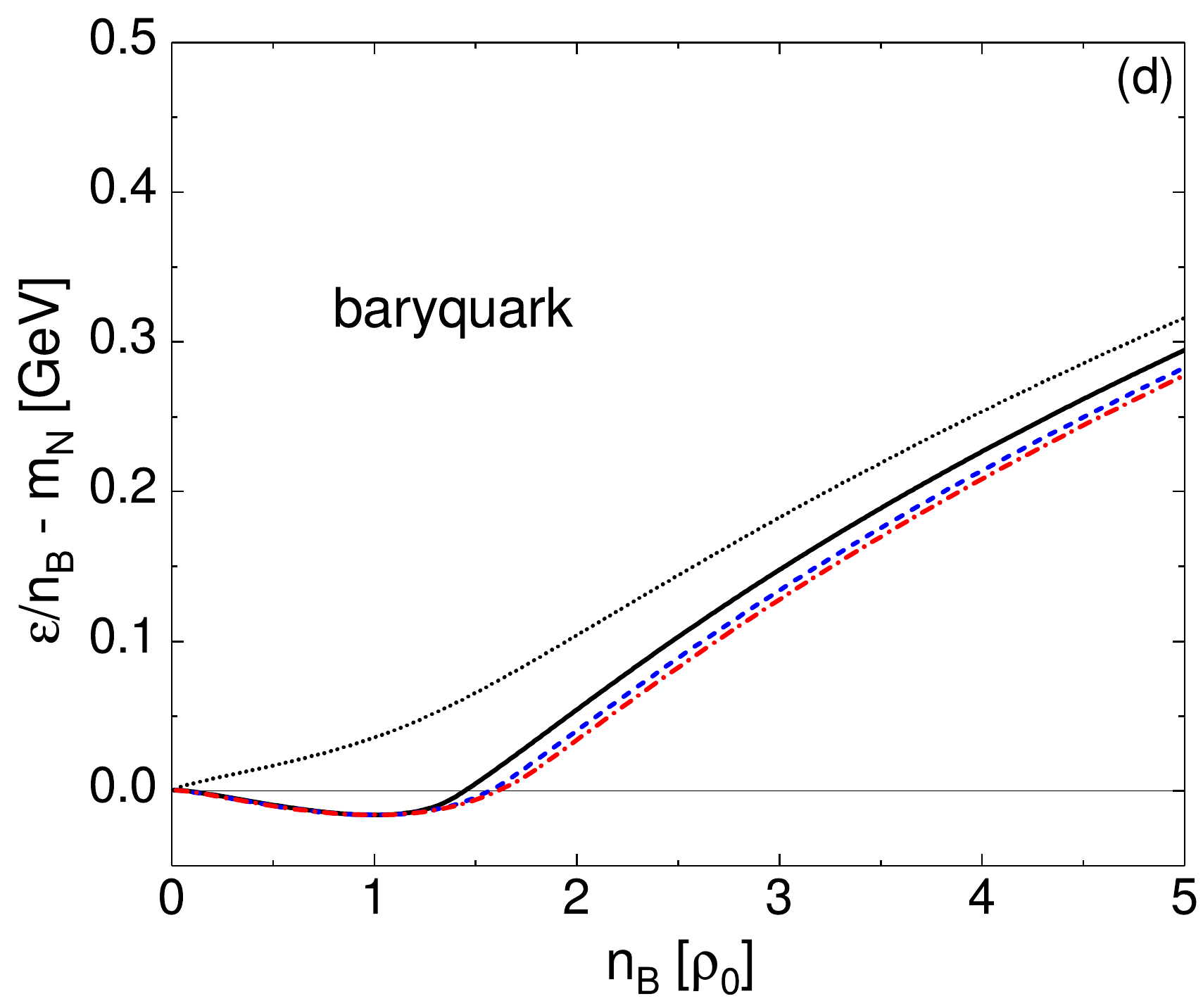}
     \includegraphics[width=.32\textwidth]{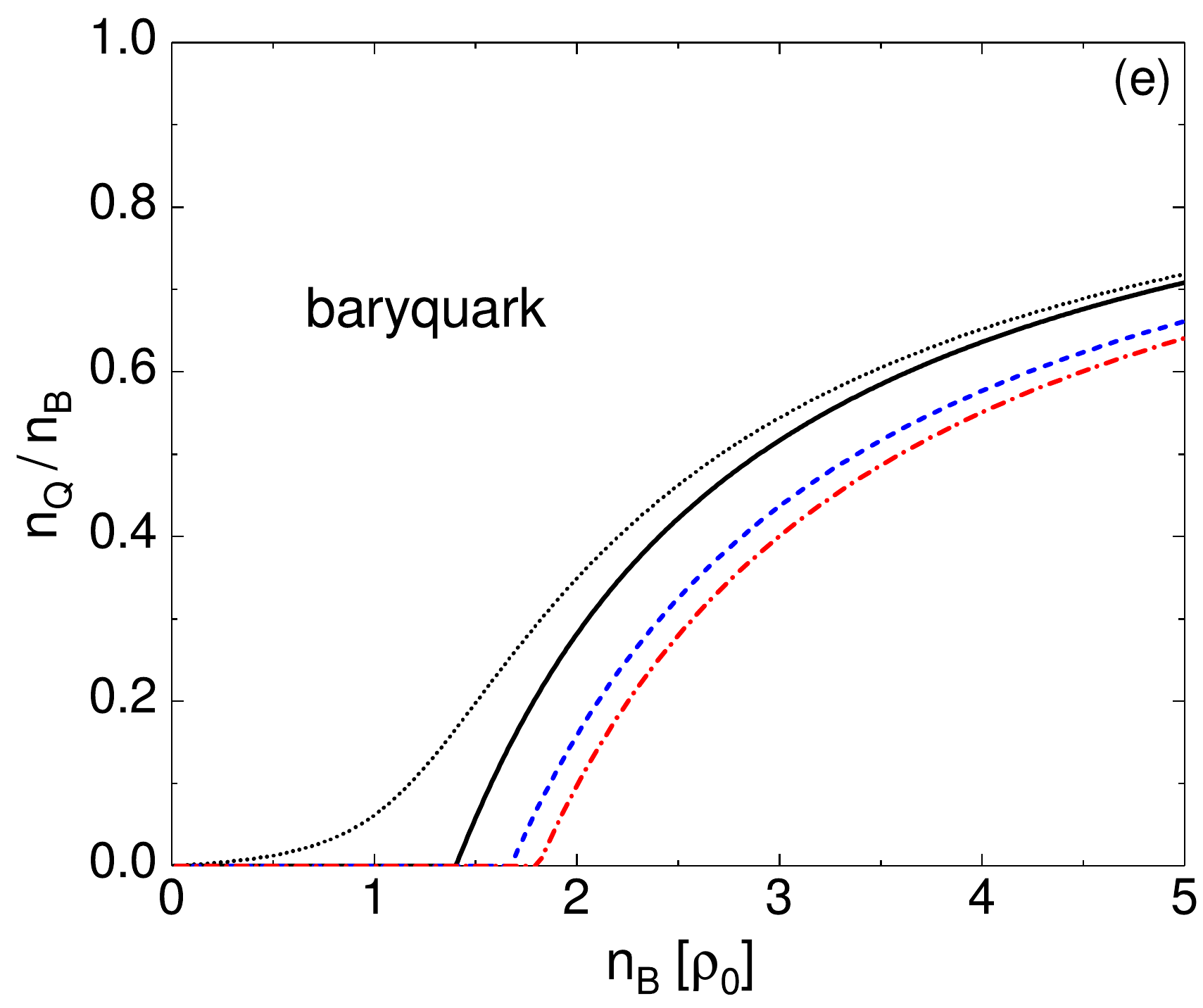}
     \includegraphics[width=.32\textwidth]{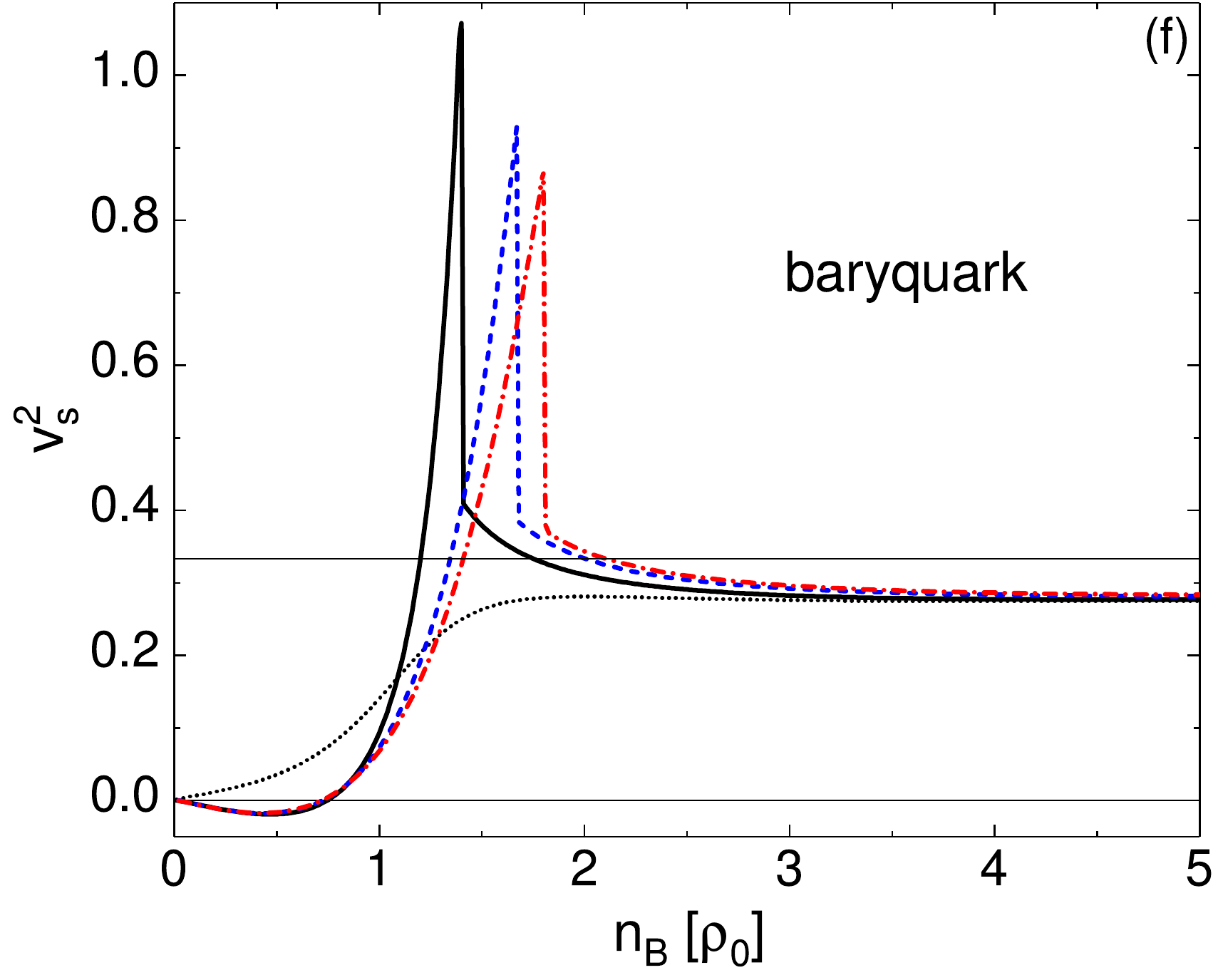}
    \caption{Density dependencies of energy per baryon (a),(d), quark fraction (b),(e), and speed of sound (c),(f) for quarkyonic matter (upper) and baryquark matter (lower). Quantum van der Waals, Carnahan-Starling, and trivirial models for nuclear interactions are presented by, respectively, solid black, dashed blue, and dash-dotted red lines. The interaction parameters $a$ and $b$ for each of the equations of state are found from the ground state properties of nuclear matter. The infrared regulator $\Lambda=0.2$~GeV is used in quarkyonic matter scenario. The thin dotted line corresponds to the vdW model without attraction (excluded volume only).
    }
    \label{fig:EoS}
\end{figure*}

The top panels in Fig.~\ref{fig:EoS} depict the baryon number density dependence of the energy per baryon number, quark fraction,  and the squared sound velocity for quarkyonic matter. Calculations are done using the three different excluded volume mechanisms: van der Waals~(solid black line), Carnahan-Starling~(dashed blue line), and trivirial~(dash-dotted red line). The thin dotted line depicts calculations in the van der Waals model without attraction~($a=0$).
We subtract the nucleon mass from $\varepsilon/n_B$ shown in the left panel for better clarity.

In all quarkyonic matter calculations we incorporate the infrared regulator $\Lambda = 0.2$~GeV described in Sec.~\ref{sec:Pauli:qy}. 
In the absence of the regulator~($\Lambda = 0$), the sound velocity exhibits acausal behavior at the onset of quark appearance, as first discussed in Ref.~\cite{Jeong:2019lhv} for the van der Waals excluded volume model without attraction. We find that neither the addition of attraction nor the modifications to the vdW excluded volume mechanism are able to cure this problem, thus, the use of the regulator is still required.
Introducing interactions in quark-nucleon or pure quark sector may thus be required, which we leave for future work~\cite{QuarkNucleonEV}.

The basic properties of the equation of state are illustrated by the density dependence of the energy per baryon, $\varepsilon/n_B - m_N$, offset by the nucleon mass $m_N = 0.938$~GeV/$c^2$~[panel (a) in Fig.~\ref{fig:EoS}].
This quantity is strictly positive, see the dotted line, and does not permit the presence of a self-bound nuclear matter if only repulsive nuclear interactions are incorporated.
In the presence of attraction, the EoS exhibits a minimum at $n_B = \rho_0$ with the value $\left. \varepsilon/n_B - m_N \right|_{n_B = \rho_0} = -16$~MeV corresponding to the binding energy of infinite matter.
This is by construction, as the QvdW parameters $a$ and $b$ in all the three models with attraction were fitted to reproduce the nuclear ground state.
As a result, the models all show very similar behavior at and below the normal nuclear density, $n_B \lesssim \rho_0$.
Dependence on the chosen excluded volume prescription becomes visible at higher densities where the transition to quarkyonic matter takes place.
At very high densities the matter becomes quark-dominated, and the behavior of the equation of state becomes insensitive to the modeling of nuclear interactions.

\begin{center}
\begin{table*}
\begin{tabular}
{ l | c | c | c | c   }
 & ~EV~  & ~QvdW~  & ~CS~  & ~TVM~ \\
\hline
\hline
 Quarkyonic matter $v_s^2$ peak density ~$n_B^{v^2_{s,\rm max}}~[\rho_0]$                & 1.43 & 1.47 & 1.76 & 1.89\\
 Quarkyonic matter $v_s^2$ peak chemical potential ~$\mu_B^{v^2_{s,\rm max}}~[\rm GeV]$                & 1.16 & 1.09 & 1.12 & 1.14\\
 Quarkyonic matter $v_s^2$ peak value~$v^2_{s, {\rm max}}$                     & 0.51 & 0.94 & 0.71 & 0.65 \\
 Baryquark matter $v_s^2$ peak density ~$n_B^{v^2_{s,\rm max}}~[\rho_0]$                & -- & 1.40 & 1.67 & 1.80 \\
 Baryquark matter $v_s^2$ peak chemical potential ~$\mu_B^{v^2_{s,\rm max}}~[\rm GeV]$                & -- & 1.05 & 1.10 & 1.12\\
 Baryquark matter $v_s^2$ peak value~$v^2_{s, {\rm max}}$                  & -- & 1.07 & 0.93 & 0.87 \\
\end{tabular}
\caption{\label{tab:vs2}
The location and value of the peak in the sound velocity $v_s^2$ in quarkyonic~(top three rows) and baryquark~(bottom three rows) matter computed within excluded volume, quantum van der Waals, Carnahan-Starling, and trivirial models for nuclear interactions.
}
\end{table*}
\end{center}

To elucidate the influence of nuclear attraction and density-dependent excluded volume repulsion on the quark-nucleon transition let us analyze the behavior of differential observables such as quark fraction and sound velocity.
Calculation results for the quark fraction $n_Q/n_B$ [panel (b) in Fig.~\ref{fig:EoS}] indicate that the matter is predominantly nucleonic at $n_N \lesssim 1.5 \rho_0$, while a sizable fraction of quarks appears at larger densities.
Without the infrared regulator~($\Lambda = 0$) the appearance of quarks would be sudden and show a kink structure in the density dependence of $n_Q/n_B$.
In the presence of a non-zero $\Lambda$, however, the kink is smoothed out, as seen in the figure. 
The baryon density at which the quarks start to appear depends on the modeling of nuclear interactions. 
In the vdW excluded volume model without attraction, the quarks start to appear already from around $n_B \sim \rho_0$.
The appearance of quarks is shifted to larger densities once the attractive nucleon mean-field is incorporated. This is also the case when we replace the vdW excluded volume prescription by CS or TVM, both of which yield a softer nuclear equation of state at $n_B \gtrsim \rho_0$.
For all three EV prescriptions considered, the quarkyonic matter regime sets in at the transition density of around $n_B^{\rm tr} \sim 1.5-1.9 \rho_0$.
These $n_B^{\rm tr}$ values are a direct prediction of our model which does not depend on any free parameters, as the nuclear interaction parameters have been fixed to reproduce the basic properties of nuclear matter.\footnote{The remaining model parameter -- the regulator $\Lambda$ -- smoothens out the transition but does nor appreciably influence the transition density $n_B^{\rm tr}$.}

We do observe that the regulator $\Lambda$ leads to an artefact where quarks appear at very low densities, $n_B \ll \rho_0$, consistent with the earlier findings~\cite{Sen:2020qcd} made for the purely excluded volume model.
At the same time, we find that the introduction of attraction makes this artefact less sizable so that it has virtually negligible impact on the equation of state. Thus, we do not introduce any density dependence into $\Lambda$.

A sensitive diagnostic of the dense matter equation of state is the (squared) sound velocity, which at zero temperature probes the second derivative of the energy density with respect to baryon density,
\eq{
v_s^2=\frac{n_B}{\mu_B}\frac{d \mu_B}{d n_B}=\frac{n_B}{\mu_B}\frac{d^2 \varepsilon}{d n_B^2}~.
}

The transition to quarkyonic matter regime is typically accompanied by a peak in the sound velocity~\cite{McLerran:2018hbz}, which may exceed the conformal limit of $v_s^2 = 1/3$.
We do observe this behavior in all our calculations, with the baryon density $n_B^{v^2_{s,\rm max}}$ at the sound velocity peak serving as a proxy for characterizing the quarkyonic matter transition density $n_B^{\rm tr}$.
The $n_B^{v^2_{s,\rm max}}$ values are shown in Table~\ref{tab:vs2}.
In all cases the maximum sound velocity values $v^2_{s,\rm max}$ lie above the conformal limit, although the situation might change if a different~(larger) value of the regulator $\Lambda$ is chosen instead.
By comparing the calculations with~(solid black lines) and without~(dotted black lines) attractive nuclear interactions, we can conclude that the presence of nuclear attraction and the associated ground state makes the peak in the sound velocity considerably sharper.
Density-dependent excluded volumes do not change the picture qualitatively but can change the location and the sharpness of the peak: the peak is shifted to larger baryon densities in CS and TVM models, and becomes broader.

The results for baryquark matter are depicted in the lower panels in Fig.~\ref{fig:EoS}.
These calculations are performed without infrared regulator, i.e. $\Lambda = 0$, thus the model has no free parameters.
We did not introduce the regulator here because, in contrast to quarkyonic matter, the excluded volume baryquark matter does not exhibit singular behavior in the sound velocity as detailed in~\cite{Koch:2022act}.
Baryquark matter has the following features in its equation of state that distinguish it from quarkyonic matter:
\begin{itemize}
    \item The quark onset is sudden and characterized by a kink in the density evolution of the quark fraction. The resulting sound velocity is finite, although it does exhibit a discontinuous drop at the quark onset. This behavior can be smoothed out by applying the regulator although its reintroduction does not appear to otherwise be required.
    \item In the absence of attraction, quarks appear much earlier and in a gradual fashion. 
    Both the quark fraction and the sound velocity are continuous and the latter shows no pronounced peak.
    The attraction thus plays a more essential role in baryquark matter compared to quarkyonic matter.
    \item Energy density at fixed baryon density is lower in baryquark matter configuration compared to quarkyonic matter, in all the considered setups. Therefore, baryquark matter is energetically favorable in these setups, extending the similar conclusion of Ref.~\cite{Koch:2022act} obtained originally for the bare excluded volume model. 
    One can also note that the sound velocity peak tends to emerge slightly earlier in baryquark matter~(see Table~\ref{tab:vs2}).
\end{itemize}

The sound velocity is causal in all cases except the vdW excluded volume model, where it exceeds unity by less than 10\% in a small baryon density interval near the quark onset.
Thus, the implementation of excluded volume prescription beyond van der Waals appears to play essential role in the modeling of baryquark matter.

Overall, the resulting equations of state turn out to be quite similar between quarkyonic and baryquark matter and it seems that the details of the implementation of the Pauli principle between nucleons and quarks has only moderate effect on thermodynamics.
This is further illustrated in Appendix~\ref{sec-uniform}, where we show the results for
uniform mixture of baryons and quarks which can be considered an intermediate case between quarkyonic and baryquark configurations.

\section{Summary}
\label{sec:summary}

We have extended the excluded volume model for quarkyonic matter by adding attractive nuclear interactions, giving rise to the quantum van der Waals model for quarkyonic matter. 
We have also explored variations on the excluded volume mechanism by incorporating Carnahan-Starling and trivirial model prescriptions.
These generalizations have allowed us to incorporate the basic properties of normal nuclear matter, such as the binding energy at saturation density, into quarkyonic as well as the recently proposed baryquark matter descriptions.
We can make the following observations based on the calculations of the resulting isospin symmetric quarkyonic and baryquark matter equations of state:
\begin{itemize}
    \item The transition to quarkyonic/baryquark matter regime takes place at a baryon density of $n_B^{\rm tr} \sim 1.5-2 \rho_0$, the exact value depending on the chosen excluded volume prescription. In contrast to earlier works, the obtained transition density does not depend on any free parameters here but is a direct prediction stemming from the incorporation of the empirical properties of normal nuclear matter. The obtained estimates suggest that quarkyonic matter can be reachable in intermediate energy heavy-ion collisions~\cite{Lovato:2022vgq,Sorensen:2023zkk}.

    \item We confirm the findings of Ref.~\cite{Jeong:2019lhv} that quarkyonic matter with the excluded volume mechanism requires the introduction of a regulator to avoid the singular behavior in the sound velocity. This conclusion does not change with the addition of attractive nuclear interactions or modifications to the excluded volume mechanism, although the presence of attraction mitigates to some extent the artifact of early quark appearance induced by the regulator. At the same time, no singular behavior in $v_s^2$ is observed for baryquark matter even in the absence of the regulator.

    \item The transition to quarkyonic matter is qualitatively similar for all three excluded volume prescriptions considered in this work.
    This is particularly notable for the trivirial (TVM) model, which has no limiting nucleon density and, in contrast to the vdW or CS excluded volume models, does not induce a singular relation between nucleon chemical potential and nucleon density at any finite 
    value of nucleon density. 
    It thus appears that the presence of such a singularity is not essential to dynamically generate the momentum shell structure. 
    Instead, the relevant property of all the different excluded volume frameworks appears to be the fact that they all modify the free gas relation between nucleon's Fermi momentum and number density~[Eq.~\eqref{eq:nNvdw}].

    \item In line with the findings of Ref.~\cite{Koch:2022act}, we observe that baryquark matter configuration is energetically favorable to quarkyonic matter in all the considered setups. At the same time, the resulting equations of state are very similar, and it would be challenging to distinguish the two scenarios based on thermodynamics alone. Other considerations, such as the introduction of quark-nucleon interactions~\cite{QuarkNucleonEV}, explicit treatment of quark-hadron duality~\cite{Fukushima:2020cmk,Kojo:2021ugu,Fujimoto:2023ioi}, or transport properties of dense matter that are sensitive to the Fermi surface physics, can thus be helpful.
    
\end{itemize}

There is a number of topics that can be studied in future works based on the present study.
A natural extension is to consider neutron star quarkyonic matter~\cite{McLerran:2018hbz} in addition to the isospin-symmetric matter studied here.
Differences between $pp$, $pn$, and $nn$ interactions can be addressed within the multi-component quantum van der Waals equation~\cite{Vovchenko:2017zpj}, with the isospin-dependent interaction parameters constrained by the symmetry energy and its slope~\cite{Poberezhnyuk:2018mwt}.
One can then explore how the properties of the quarkyonic matter transition depend on the isospin asymmetry, and the resulting implications for neutron stars.

The modeling of nuclear interactions can also be improved. 
As one can see from Table~\ref{tab:NM}, the models explored here overestimate the value of the nuclear incompressibility factor considerably. 
It is thus important to explore how the equation of state can be affected if the modeling of nuclear interactions is improved further.
In particular, we do observe that the quark onset density $n^{\rm tr}_B$ tends to be anticorrelated to the resulting incompressibility factor~(Table~\ref{tab:NM}), thus it may be that the further softening of the nuclear equation of state would push $n^{\rm tr}_B$ to higher values beyond $2\rho_0$.

Finally, the formalism could be extended to finite temperatures~\cite{Stoecker:1980uk,Sen:2020peq,Duarte:2021tsx}, enabling applications to heavy-ion collisions and neutron star mergers.

\begin{acknowledgments}

\emph{Acknowledgments.} 
The authors thank Kie Sang Jeong 
for the detailed reading of the manuscript and useful suggestions.
We also thank Yuki Fujimoto, Volker Koch,  and Jan Steinheimer for fruitful comments and discussions.
R.V.P. acknowledges the support from 
the Philipp Schwartz Initiative of the Alexander von Humboldt Foundation. 
H.St. appreciates the Judah M. Eisenberg Professur Laureatus of the Walter Greiner Gesellschaft/F\"orderverein f\"ur physikalische Grundlagenforschung Frankfurt, and the Fachbereich Physik at Goethe Universit\"at.

\end{acknowledgments}

\appendix

~\\

\section{Uniform mixture of baryons and quarks}
\label{sec-uniform}

\begin{figure}[h!]
    \includegraphics[width=.45\textwidth]{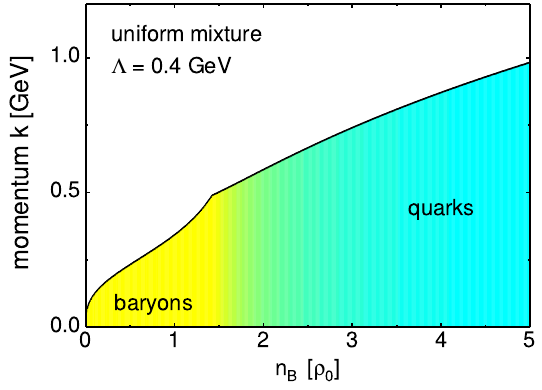}
    \caption{Same as in Fig.~\ref{fig:shell-structures} but for uniform 
    mixture of baryons and quarks with $\Lambda = 0.4$~GeV. 
    }
    \label{fig:uniform-shell-structure}
\end{figure}

One simple configuration for the realization of the Pauli exclusion principle is the uniform (homogeneous) distribution of confined and deconfined quarks in momentum space,
corresponding to a constant density of states modification factor $0 \leq \rho \leq 1$:
\eq{\label{eq:uni}
\rho_Q(q) & = \rho,\\
\rho_N(k) & = 1-\rho.
}
There is no momentum shell structure in this case.
Instead, each quark state has the same probability $\rho$ to be occupied by a deconfined quark, and probability $1-\rho$ otherwise.
Thus, the factor $\rho$ simply corresponds to the quark fraction $n_Q/n_B$.
This is shown in Fig.~\ref{fig:uniform-shell-structure} which depicts the density dependence of momentum space distribution of uniform 
mixture (for $\Lambda = 0.4$~GeV).
The change of the color map from yellow to cyan reflects the increasing quark fraction $n_Q/n_B$.

\begin{figure}
    \includegraphics[width=.45\textwidth]{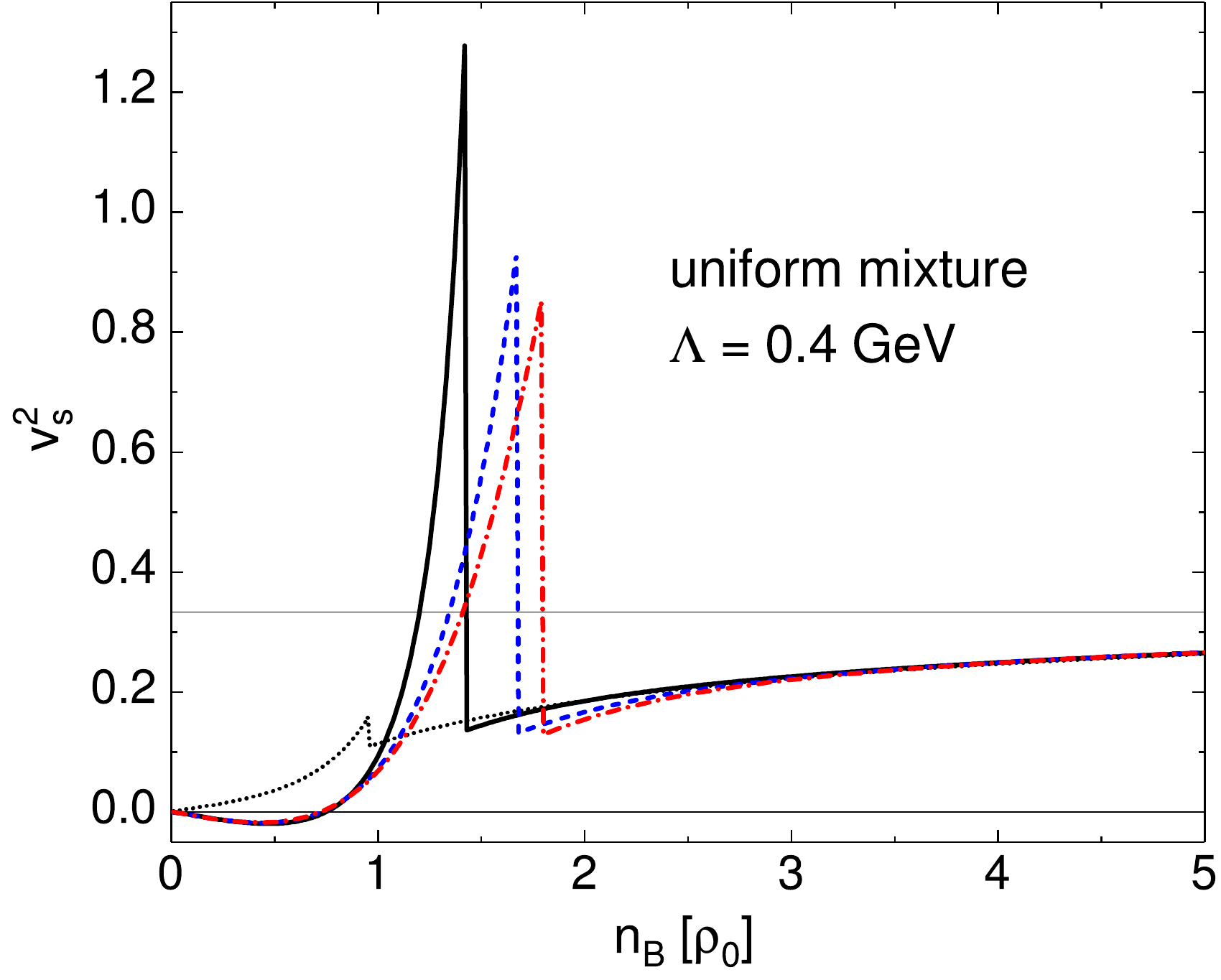}
    \caption{Same as in Fig.~\ref{fig:EoS}(c),(f)  but for uniform mixture of baryons and quarks. An infrared regulator with $\Lambda=0.4$~GeV is used to limit the peak in the speed of sound.}
    \label{fig:uniform}
\end{figure}

The equation of state is shown in Fig.~\ref{fig:uniform} depicting the speed of sounds as a function of $n_B$.
The quark onset density $n_B^{\rm tr}$ is similar to that of baryquark matter~(1.4-1.8~$\rho_0$).
Similar to quarkyonic matter, the sound velocity is singular at $n_B^{\rm tr}$, therefore, the regulator $\Lambda$ is required, which we take here to be $\Lambda=0.4$~GeV.
Similar to baryquark matter, the sound velocity is discontinuous at $n_B^{\rm tr}$.

Note that in the present paper we consider deconfined quarks in the ideal gas framework.
However, the 
assumption of the common Fermi surface for confined and deconfinded quarks implies interactions between them.
Thus, in future, a more elaborate description should be applied to test the uniform 
mixture consistently. 

\bibliography{main}

\end{document}